\documentclass[twocolumn,final,aps,prb,floatfix,amsmath,amssymb,showpacs]{revtex4}
\usepackage{graphicx,amssymb}
\usepackage{natmove}
\DeclareGraphicsExtensions{.pdf,.eps,.ps,.pstex}
\graphicspath{{./}{./Figures/}}%
\usepackage{prettyref}%
	\newcommand{\pref}[1]{\prettyref{#1}}%
	\newrefformat{fig}{Fig.\ \ref{#1}}%
	\newrefformat{tab}{Table \ref{#1}}%
	\newrefformat{sec}{Sec.\ \ref{#1}}%
	\newrefformat{eq}{Eq.\ (\ref{#1})}%
\newcommand{\perR}[1]{\ensuremath{#1\sqrt{3}\!\times\!#1\sqrt{3}\,R30^{\circ}}}

\newcommand{\per}[1]{\ensuremath{#1\times #1}}
\newcommand{\DzG}{\ensuremath{\Delta z_{\mathrm{G}}}}
\newcommand{\Ebnd}{\ensuremath{E_{\mathrm{int}}}}
\newcommand{\Estr}{\ensuremath{E_{\mathrm{str}}}}
\newcommand{\Edef}{\ensuremath{E_{\mathrm{def}}}}
\newcommand{\Estk}{\ensuremath{E_{\mathrm{stick}}}}
\newcommand{\ECSi}{\ensuremath{E_{\mathrm{C-Si}}}}
\newcommand{\thp}{\ensuremath{\theta_{\mathrm{p}}}}
\newcommand{\thsp}{\ensuremath{\theta_{\sigma\pi}}}
\newcommand{\adeg}[1]{\ensuremath{#1^{\circ}}}

\newcommand{\eV}{\ensuremath{\mathrm{eV}}}
\usepackage{xcolor}

\begin{document}


\title{%
Carbon Rehybridization at the Graphene/SiC(0001) Interface: Effect on Stability and Atomic-Scale Corrugation
}
\date{\today}
\author{Gabriele Sclauzero}
\email{gabriele.sclauzero@epfl.ch}
\author{Alfredo Pasquarello}
\affiliation{
Chaire de Simulation \`a l'Echelle Atomique (CSEA), Ecole Polytechnique F\'ed\'erale de Lausanne (EPFL), CH-1015 Lausanne, Switzerland}
\begin{abstract}
  We address the energetic stability of the graphene/SiC(0001) interface and the associated binding mechanism by studying a series of low-strain commensurate interface structures within a density functional scheme.
  Among the structures with negligible strain, the $6\sqrt{3}\!\times\!6\sqrt{3}\,R30^{\circ}$ SiC periodicity shows the lowest interface energy, providing a rationale for its frequent experimental observation.
  The interface stability is driven by the enhanced local reactivity of the substrate-bonded graphene atoms undergoing $sp^2$-to-$sp^3$ rehybridization (pyramidalization).
  By this mechanism, relaxed structures of higher stability exhibit more pronounced graphene corrugations at the atomic scale. 
\end{abstract}

\maketitle


Epitaxially grown graphene on SiC provides a promising solution for the realization of high-performance carbon-based electronic devices \cite{Berger2004JPCB,Berger2006Science,Zhou2007NatMat,Lin2010Science,Sprinkle2010NatNano}.
On the Si-face of SiC, the interface between the epitaxial graphene overlayers and the SiC substrate consists of a thin carbon buffer layer \cite{Forbeaux1998PRB,Mattausch2007PRL,Varchon2007PRL,Emtsev2008PRB} which has been characterized through several experimental techniques \cite{Forbeaux1998PRB,Berger2004JPCB,Chen2005SS,Riedl2007PRB,Rutter2007PRB,Emtsev2008PRB,Zhou2007NatMat}.
Angle-resolved photoemission spectroscopy data show the presence of well-developed $\sigma$ bands and are thus consistent with a buffer layer having the same topology as graphene \cite{Emtsev2008PRB}.
However, the absence of the free-standing graphene Dirac cones in the $\pi$ bands points to the occurrence of covalent bonding between the buffer layer and SiC(0001) \cite{Emtsev2008PRB,Mattausch2007PRL,Varchon2007PRL}, as also corroborated by x-ray photoemission data \cite{Chen2005SS,Emtsev2008PRB}.
A moir\'e pattern of \perR{6} SiC periodicity usually arises in electron diffraction \cite{Berger2004JPCB,Chen2005SS,Riedl2007PRB,Emtsev2008PRB} and in scanning tunneling measurements \cite{Riedl2007PRB,Cervenka2010PSSa} due to the lattice constant mismatch between graphene and SiC.
The observed moir\'e is commonly understood as a coincidence lattice between a \perR{6} SiC surface reconstruction and a \per{13} graphene sheet \cite{Forbeaux1998PRB,Kim2008PRL,Varchon2008PRB,Emtsev2008PRB}.
However, several other SiC periodicities, such as \perR{2}, \per{4}, \per{5}, \ldots, are compatible with an almost strain-free graphene/SiC interface \cite{Sclauzero2012DRM,Pankratov2010PRB,Hass2008JPCM}, though only few of them have been found in experiments \cite{Riedl2007PRB,Hass2008JPCM}.
A further indication that the existence of a low-strain coincidence lattice is not a sufficient condition for its experimental realization can be inferred from the observation of a \adeg{2.2}-rotated commensurate graphene phase on the C face \cite{Hass2008PRL}, which does not have a counterpart on the Si face.
Hence, legitimate questions arise about the underlying physical mechanisms which drive the occurrence of specific structures at the graphene/SiC(0001) interface. 

In this Rapid Communication, we address the energetics of the graphene/SiC(0001) interface by considering a series of model structures differing by the SiC periodicity, the strain in the graphene layer, and the rotation angle between the graphene and SiC lattices.
Our study shows that the experimentally observed \perR{6} periodicity is favored over other low-strain commensurate structures by its higher binding energy. 
An analysis of the binding energies in terms of C-Si covalent bond energies reveals that the driving mechanism leading to interface stability is dominated by the $sp^2$-to-$sp^3$ rehybridization undergone by the substrate-bonded graphene atoms.
As a consequence, a clear connection is established between the binding energy and the atomic-scale corrugations of the graphene layer.

In our density functional theory calculations, the exchange-correlation energy is described within a generalized gradient approximation (GGA) \cite{perdew1996}.
We also checked that the description of van der Waals interactions beyond the GGA level \cite{Grimme2006JCC,Barone2009JCC} does not change our conclusions.
The valence wave functions and the electron charge are expanded in plane wave basis sets defined by kinetic energy cutoffs of 35 and 350 Ry \cite{Laasonen1993}, respectively.
The core-valence interaction is described through ultrasoft pseudopotentials \cite{Laasonen1993}.
Our slab geometries comprise one graphene layer and five bilayers of SiC, of which the bottom C atoms are saturated with H atoms.
The slab is separated from its periodic replicas by vacuum regions of 30 \AA. 
For each interface system the sampling of the Brillouin zone is at least equivalent to a $10\times 10$ $k$-point mesh in the primitive cell of graphene, which ensures errors lower than 0.01 eV on the total energy per surface Si atom.
All atoms but those in the bottom two SiC bilayers are allowed to relax.
We used the \textsc{Quantum ESPRESSO} package \cite{Giannozzi2009}.

\begin{table}[b]
  \setlength{\tabcolsep}{4pt}
  \caption{Energetics and structural properties of the relaxed graphene/SiC(0001) interfaces: graphene/SiC crystallographic rotation angle $\alpha$, strain $s$ on the graphene sheet, interface energy per surface Si atom \Ebnd, and vertical spread of the graphene C atoms \DzG.
  \Ebnd\ is decomposed in strain energy \Estr, deformation energy \Edef, and sticking energy \Estk\ of the graphene layer. 
  The interfaces are separated in three groups according to strain: $|s|\leqslant 0.2\%$, $s>0.2\%$, and $s<-0.2\%$.
  Energies are in \eV\ and \DzG\ in \AA.}
  \label{tab:results}
  \begin{tabular}[c]{lrrccccc}
    \hline\hline
    &\multicolumn{1}{c}{$\alpha$} &\multicolumn{1}{c}{$s$} & \Ebnd & \Estr & \Edef & \Estk &  \DzG \\
    \hline
R6           & \adeg{30.0} & $-$0.02\%   &  $-$0.32  & $<0.001$&   0.77  &  $-$1.09  &   0.90 \\
N9           & \adeg{27.5} & $-$0.12\%   &  $-$0.31  & $<0.001$&   0.79  &  $-$1.10  &   0.98 \\
N5           & \adeg{16.2} &  0.14\%     &  $-$0.28  & $<0.001$&   0.78  &  $-$1.07  &   0.86 \\
R6$^{\prime}$&  \adeg{2.2} & $-$0.02\%   &  $-$0.19  & $<0.001$&   0.37  &  $-$0.56  &   0.77 \\
N4           &  \adeg{0.0} &  0.06\%     &  $-$0.19  & $<0.001$&   0.36  &  $-$0.55  &   0.74 \smallskip\\
R1    & \adeg{30.0} &  8.31\%   & \phantom{$-$}0.24  &   1.010 &   0.50  &  $-$1.27  &   0.28 \\
R4$^{\prime\prime}$& \adeg{24.2} &  1.42\%   &  $-$0.34  &   0.040 &   0.73  &  $-$1.10  &   0.64 \\%
N7   & \adeg{23.4} &  0.43\%   &  $-$0.32  &   0.004 &   0.76  &  $-$1.08  &   0.68 \smallskip\\
R2   &  \adeg{6.6} & $-$0.60\%   &  $-$0.16  &   0.008 &   0.51  &  $-$0.68  &   0.88  \\
R4$^{\prime}$  &  \adeg{6.6} & $-$0.60\%   &  $-$0.18  &   0.008 &   0.70  &  $-$0.89  &   0.83 \\
    \hline\hline
  \end{tabular}
\end{table}

Focusing on \per{m} and \perR{m} SiC periodicities (denoted as N$m$ and R$m$, respectively), we search for commensurate interface structures of tractable size in which the graphene sheet is subject to limited strain, either tensile or compressive \cite{Sclauzero2012DRM}.
This procedure results in a series of interface structures, each of them characterized by an integer $m$ and a rotation angle between the graphene and SiC lattices (\pref{tab:results}).
Structures with the same SiC periodicity but with different rotation angle are distinguished by primes.
The largest SiC periodicity considered here corresponds to the experimentally observed \perR{6}, requiring up to 1526 atoms in the simulation cell.
For all retained structures, the graphene strain is below $1.5\%$ (cf.\ \pref{tab:results}) except for R1 ($8.3\%$). 
The latter presents an unrealistic strain level, but has been included here for comparison with previous work.\cite{Mattausch2007PRL,Varchon2007PRL} 
For each structure, we also report in \pref{tab:results} the calculated interface energy per surface Si atom \Ebnd, referred to ideal, unstrained graphene and unreconstructed SiC(0001).
Apart from the R1 interface, all interface structures are energetically stable ($\Ebnd<0$). 

\begin{figure}[tb]
  \includegraphics[width=0.9\columnwidth]{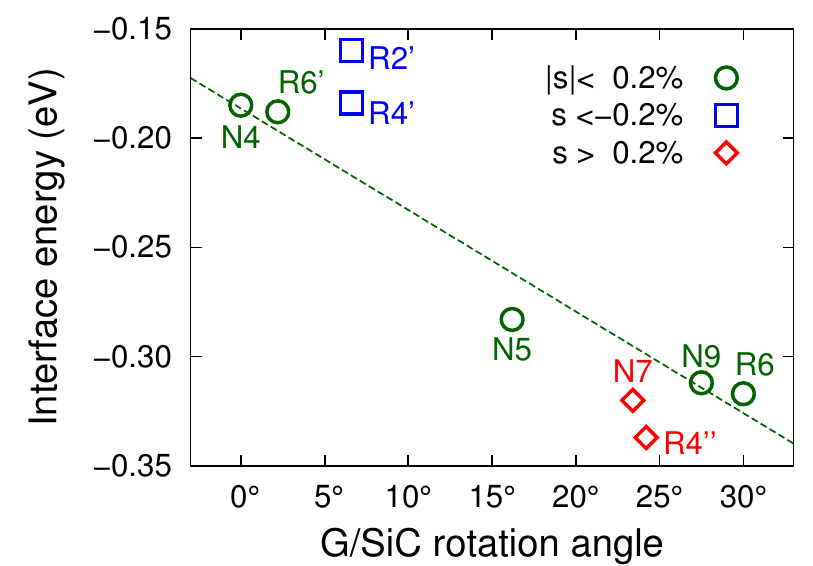}
  \caption{(Color online) Interface energy \Ebnd\ vs graphene/SiC rotation angle $\alpha$.
  The structures are distinguished by the strain $s$ in the graphene layer: very weak strain with $|s|\leqslant 0.2\%$ (circles), tensile strain with $s>0.2\%$ (diamonds), and compressive strain with $s<-0.2\%$ (squares). The dashed line is a linear fit of \Ebnd\ for the structures with $|s|\leqslant 0.2\%$.}
  \label{fig:EbndAngle}
\end{figure}

As shown in \pref{fig:EbndAngle}, the energies of interfaces with strain lower than $0.2\%$ decrease monotonically with increasing graphene/SiC rotation angle.
The lowest energy is found for the \perR{6} periodicity (R6), providing an energetic argument for its frequent experimental observation.
Semi-empirical corrections accounting for the van der Waals interactions \cite{Grimme2006JCC} just shift the interface energies by a constant amount and thus do not affect this comparison.
With respect to the low-strain interfaces with close rotation angles, the structures with higher tensile and compressive strain yield lower and higher interface energies, respectively.
This behavior is consistent with the dependence of the chemical reactivity of graphene on strain \cite{deAndres2008APL}, leading in particular to a very stable R4$^{\prime\prime}$ interface. 
However, to the best of our knowledge, this periodicity has not been observed experimentally, suggesting that such levels of graphene strain cannot be sustained during growth.

To investigate the factors playing a role in the interface stability, the interface energy has been decomposed as
\begin{equation}
  \Ebnd = \Estr + \Edef + \Estk,
  \label{eq:enedec}
\end{equation}
where \Estr\ is the energy cost to strain ideal graphene without deformations, \Edef\ is the energy needed to deform the strained graphene layer into the final atomic configuration, and \Estk\ is the energy gained by sticking the so-obtained graphene layer onto SiC.
In this decomposition, the deformation energy of the SiC substrate is included in \Estk\ and is not singled out because its contribution is found to be negligible ($<0.02$ eV per surface Si atom).
The calculated values of the three components of the binding energy in \pref{eq:enedec} are given in \pref{tab:results}. 
Focusing first on interfaces with strain lower than $0.2\%$, we notice that the strain energy is much smaller than the interface energy and thus does not influence the stability. 
Furthermore, one observes that \Ebnd\ results from a compensation effect between the cost of \Edef\ and the gain of \Estk, with \Estk\ roughly scaling with \Edef. 
In turn, this directly connects the interface energy with the deformation energy, resulting in higher stabilities for larger degrees of graphene deformation.
Overall, these observations qualitatively also hold for the interfaces with larger strains except for R1, in which the considerable strain energy makes the structure unstable. 

\begin{figure}[t]
  \includegraphics[width=0.95\columnwidth]{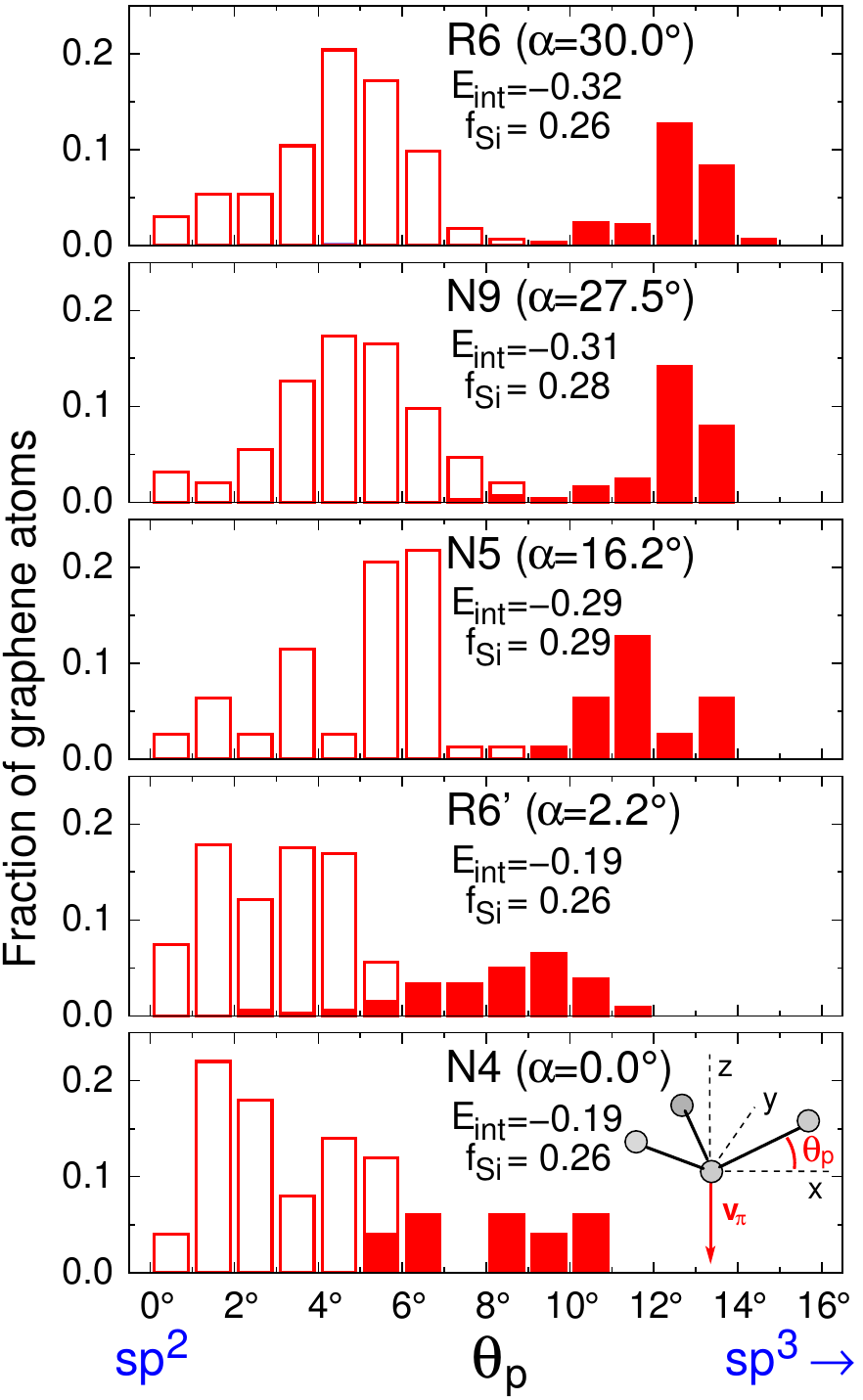}
  \caption{(Color online) Distribution of the pyramidalization angle \thp\ of the C atoms in the buffer layer for the interface structures with $|s|\leqslant0.2\%$. 
  The solid bars refer to Si-bonded graphene C atoms identified by a cutoff distance of 2.3 \AA, while the open bars refer to non-bonded C atoms.
  The graphene/SiC rotation angle $\alpha$, the interface energy \Ebnd, and the fraction $f_{\mathrm{Si}}$ of Si-bonded graphene C atoms are indicated for each structure.
  In the bottom panel, the inset illustrates the $\pi$-axial vector $\textbf{v}_{\pi}$ and the pyramidalization angle $\theta_p$.}
  \label{fig:pyram}
\end{figure}

To better understand their relation with the interface energy, we analyze in more detail the nature of the deformations undergone by graphene.
It has been suggested that the structural properties of the interface could be driven by dangling bond saturation of the surface Si atoms through bond formation with the graphene C atoms \cite{Emtsev2008PRB,Sclauzero2012DRM}.
Since the ideal moir\'e pattern also presents some surface Si atoms which do not have any C atom directly atop \cite{Emtsev2008PRB,Cervenka2010PSSa,Sclauzero2012DRM}, one could then argue that deformations are mainly induced by lateral adjustments of the C atoms.
However, it has been shown that the interface energy as described in a bond energy picture does not undergo significant variations when comparing the case of undistorted graphene with that of the fully relaxed buffer layer \cite{Sclauzero2012DRM}.
Furthermore, when focusing on interface structures with $|s|<0.2\%$ (\pref{tab:results}), one observes that the deformation energy approximately scales with the vertical spread of the graphene C atoms \DzG\ \cite{Sclauzero2011MEE}, suggesting that the main contributions come from out-of-plane deformations rather than from lateral displacements.

In view of these considerations, we further analyze the deformations by focusing on the \emph{rehybridization} undergone by the graphene C atoms.
Rehybridization effects have already been pointed out to explain the pinning of graphene to a metallic substrate upon deposition of adsorbates \cite{Feibelman2008PRB}.
To this purpose, we consider the pyramidalization angle $\thp$ (Ref.~\onlinecite{Haddon1988ACR}) formed by each of the C atoms in the buffer layer. 
As shown in the inset in \pref{fig:pyram}, the definition of $\thp$ is based on the construction of the $\pi$-orbital axis vector $\textbf{v}_{\pi}$, which forms equal angles \thsp\ with the three $\sigma$-bonds connecting the central C atom to its graphene neighbors.
The pyramidalization angle is then obtained as $\thp=\thsp-\adeg{90}$.
It vanishes for a C atom with $sp^2$ hybridization as in the ideal honeycomb structure, and takes the value of $\adeg{19.47}$ for a C atom in the tetrahedral $sp^3$ configuration.
The distributions of the calculated pyramidalization angles of the C atoms in the buffer layer are shown in \pref{fig:pyram} for the interface structures with $|s|<0.2\%$.
For all the structures considered here, the angle distribution is bimodal.
To elucidate this property, we separate the graphene C atoms in two groups by identifying those forming bonds to surface Si atoms (\pref{fig:pyram}).
This decomposition shows that graphene C atoms forming a C-Si bond are responsible for the peak at high pyramidalization angles, while the non-bonded C atoms give rise to the feature at lower \thp.

\begin{figure}[b]
  \includegraphics[width=0.95\columnwidth]{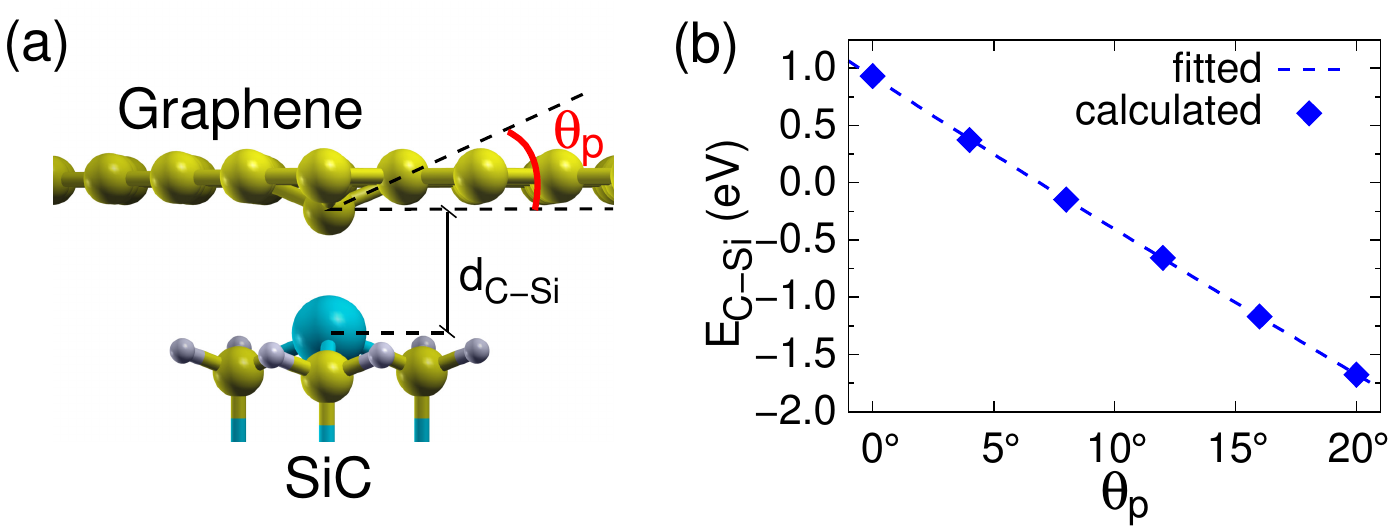}
  \caption{(Color online) (a) Atomistic configuration used to compute \ECSi\ as a function of the pyramidalization angle \thp\ [only the top of the SiC pyramid construction (Ref.~\onlinecite{Sclauzero2012DRM}) is shown].
  (b) Calculated \ECSi\ vs \thp\ and fit according to \pref{eq:ECSi} (dashed line).
  We used a fixed C-Si distance $d_{\mathrm{C-Si}}$ of 2.05 \AA, corresponding to typical C-Si bond distances at the graphene/SiC(0001) interface.}
  \label{fig:fitECSi}
\end{figure}

The fraction of Si-bonded graphene atoms does not vary significantly among the considered interface structures (\pref{fig:pyram}) and cannot explain the differences in the calculated interface energies, as noted above.
In this perspective, it is important to observe that the peak at high pyramidalization angles shifts significantly toward higher values as interfaces of increasing stability are considered.
This suggests a direct link between the local C-Si bond energy and the pyramidalization angle of the involved C atom.
Indeed, the bond energy of an individual C-Si bond can be expressed in terms of the rehybridization of the C $\pi$-orbital as \cite{Haddon1988ACR,Park2003NL}:
\begin{equation}
  \ECSi(\thp) = \epsilon_{\mathrm{gr}} + \sqrt{2} \tan{\thp} \, \epsilon_{\mathrm{s}} + 
    \sqrt{1 - 2 \tan^2{\thp}} \, \epsilon_{\mathrm{p}},
  \label{eq:ECSi}
\end{equation}
where $\epsilon_{\mathrm{gr}}$ is the energy cost for isolating a single $\pi$ orbital from the conjugated $\pi$ system of graphene, and the matrix elements $\epsilon_{\mathrm{s}}$ and $\epsilon_{\mathrm{p}}$ involve surface Si orbitals and C orbitals of $s$ and $p_z$ character, respectively.
Since $\epsilon_{\mathrm{s}} < \epsilon_{\mathrm{p}}$, C atoms with larger \thp\ are more reactive and lead to stronger C-Si bonds.

\begin{figure}[tb]
  \includegraphics[width=0.9\columnwidth]{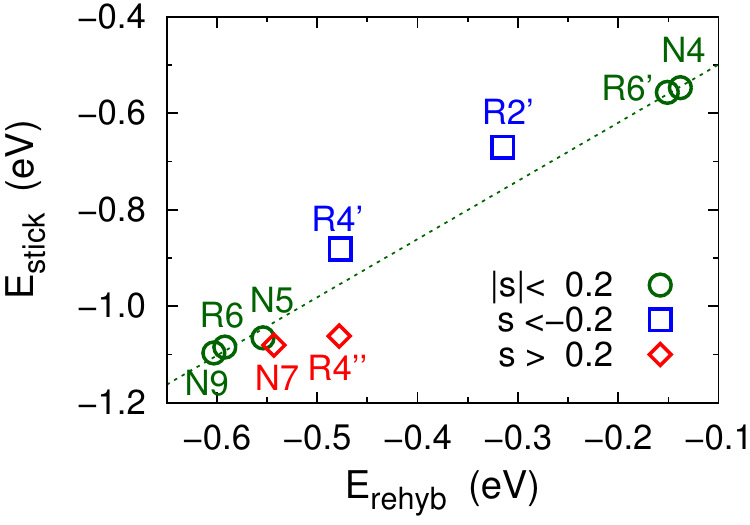}
  \caption{(Color online) Sticking energy \Estk\ of the graphene buffer layer vs model energy $E_{\textrm{rehyb}}$ accounting for the C $\pi$-orbital local rehybridizations.
  Both energies are given per surface Si atom.
  The structures are distinguished according to strain, as in \pref{fig:EbndAngle}.}
  \label{fig:EstkEp}
\end{figure}

To evaluate the bond energy variations as a function of \thp, we use an atomistic model consisting of a pyramidal structure in which a single surface Si atom is exposed to the graphene layer \cite{Sclauzero2012DRM}, as illustrated in \pref{fig:fitECSi}(a).
In the range of interest ($\thp<\adeg{20}$), the dependence on \thp\ is found to be approximately linear [\pref{fig:fitECSi}(b)], as expected from \pref{eq:ECSi}. 
The calculated \ECSi\ values are well reproduced by \pref{eq:ECSi} with $\epsilon_{\mathrm{gr}}=2.8\,\eV$, $\epsilon_{\mathrm{s}}=-5.6\,\eV$ and $\epsilon_{\mathrm{p}}=-1.9\,\eV$.
In this way, we obtain for each Si-bonded graphene C atom a bond energy $\ECSi(\thp)$ from its \thp\ value in the relaxed interface structure.
This allows us to define a model energy  $E_{\textrm{rehyb}}$ based on the sum of all these individual contributions which account for the local trigonal hybridization of the C $\pi$ orbitals.

In \pref{fig:EstkEp}, we confront $E_{\textrm{rehyb}}$ with the sticking energy \Estk\ for all structures in \pref{tab:results} but R1. 
\footnote{Here it is more appropriate to compare $E_{\textrm{rehyb}}$ with \Estk, rather than \Ebnd, because the bond energy in \pref{eq:ECSi} does not account for the local deformation energy of graphene.
However, given that the graphene deformation energy \Edef\ scales with \Estk\ for the interfaces with negligible strain, $E_{\textrm{rehyb}}$ then also correlates with \Ebnd.
The deformation energy associated to the displacements of the surface Si atoms is instead much smaller and does not play a relevant role.}
The structures with graphene strain $|s|<0.2\%$ show that the variations in \Estk\ are fully captured by the variations of $E_{\textrm{rehyb}}$.
This correlation approximately also holds when considering the structures of higher strain ($|s|>0.2\%$), indicating that the C reactivity is affected by strain effects to a lesser extent.
These considerations demonstrate that the interfacial binding mechanism is directly related to the atomic-scale corrugation of the graphene and to the associated enhancement of the chemical reactivity of the Si-bonded graphene C atoms due to their partial $sp^2$-to-$sp^3$ rehybridization.

In conclusion, we highlight the role of carbon rehybridization in the binding of the graphene layer to the SiC(0001) substrate. 
This notion carries general validity and is expected to hold for other graphene interfaces involving covalent bonding.
The pyramidalization of the carbon atoms also connects with the atomic-scale corrugation of the graphene layer and could further lead to a spatial modulation of the graphene reactivity toward adsorbates.

Partial financial support is acknowledged from the Swiss National Science Foundation (Grants No.\ 200020-119733/1 and No. 206021-128743).
We used the computational resources of CSEA-EPFL and CADMOS.
The financial support for CADMOS and the Blue Gene/P system is provided by the Canton of Geneva, Canton of Vaud, Hans Wilsdorf Foundation, Louis-Jeantet Foundation, University of Geneva, University of Lausanne and \'Ecole Polytechnique F\'ed\'erale de Lausanne.


\end{document}